\documentclass[
 reprint,
 superscriptaddress,
 amsmath,
 amssymb,
 aps,
 amsmath,
 amssymb,
 longbibliography,
 pra
]{revtex4-1}

\usepackage{graphicx}
\usepackage{dcolumn}
\usepackage{bm}
\usepackage{amsmath,amssymb,amstext,amsthm}
\usepackage{braket}
\usepackage{bbold}
\usepackage{bbm}
\usepackage{xcolor}
\usepackage[running]{lineno}
\usepackage[separate-uncertainty = true]{siunitx}
\usepackage{soul}
\usepackage{ulem}
\usepackage[hidelinks]{hyperref}
\usepackage{xfrac}
\usepackage{cuted}

\colorlet{RED}{red}
\DeclareSIUnit\angstrom{\text{\AA}}

\begin{document}
\preprint{APS/123-QED}
\setlength{\abovedisplayskip}{1pt}
\title{Design and Monte Carlo Simulation of a Phase Grating Moir\'e Neutron Interferometer to Measure the Gravitational Constant}

\author{C. Kapahi} 
\email{c3kapahi@uwaterloo.ca}
\affiliation{Institute for Quantum Computing, University of Waterloo,  Waterloo, ON, Canada, N2L3G1}
\affiliation{Department of Physics, University of Waterloo, Waterloo, ON, Canada, N2L3G1}
\author{D. Sarenac}
\affiliation{Department of Physics, University at Buffalo, State University of New York, Buffalo, New York 14260, USA}
\author{B. Heacock}
\affiliation{National Institute of Standards and Technology, Gaithersburg, Maryland 20899, USA}
\author{D. G. Cory}
\affiliation{Institute for Quantum Computing, University of Waterloo,  Waterloo, ON, Canada, N2L3G1}
\affiliation{Department of Chemistry, University of Waterloo, Waterloo, ON, Canada, N2L3G1}
\author{M. G. Huber}
\affiliation{National Institute of Standards and Technology, Gaithersburg, Maryland 20899, USA}
\author{J. W. Paster}
\affiliation{National Institute of Standards and Technology, Gaithersburg, Maryland 20899, USA}
\author{R. Serrat}
\affiliation{Institute for Quantum Computing, University of Waterloo,  Waterloo, ON, Canada, N2L3G1}
\affiliation{Centre de Formació Interdisciplinària Superior, Universitat Politècnica de Catalunya, 08028 Barcelona, Spain}
\author{D. A. Pushin}
\email{dmitry.pushin@uwaterloo.ca}
\affiliation{Institute for Quantum Computing, University of Waterloo,  Waterloo, ON, Canada, N2L3G1}
\affiliation{Department of Physics, University of Waterloo, Waterloo, ON, Canada, N2L3G1}

\date{\today}

\begin{abstract}

The gravitational constant (G) is the least precisely known fundamental constant of nature, with persistent and significant discrepancies between measurement methods.
New techniques for measuring G with systematic effects different from commonly applied pendulum methods are required.
Neutrons are convenient probes of gravitational forces as they are both massive and electrically neutral, properties that allowed a single-crystal neutron interferometer (NI) to achieve the first experimental demonstration of gravitationally induced quantum interference.
Despite this, the limitation of single-crystal NIs to monoenergetic beams significantly reduces neutron flux, making precision gravitational measurements unfeasible.
A new NI design called the phase-grating moir\'{e} interferometer (PGMI) has been shown to increase neutron flux by orders of magnitude while allowing grating separation that maintains similar interferometer area to previous NI devices.
Here, we propose and describe an experiment to measure G using the PGMI to a precision comparable to measurements from the CODATA 2022 evaluation.
A Monte Carlo model for incorporating nonlinear potentials into a PGMI is introduced.
This model is used to evaluate sources of systematic uncertainty and quantify the uncertainty in G arising from these effects.
The effect of lunar gravitation on a torsion pendulum experiment from CODATA 2022 is calculated, and the need for possible correction factors is demonstrated.
This work demonstrates that a neutron PGMI can be used to measure G to $150$ parts-per-million in the near term with the potential to achieve greater precision in future experimental designs.
\end{abstract}
\maketitle

\section{Introduction}
\label{sec:intro}

\begin{figure}
    \centering
    \includegraphics[width=\linewidth]{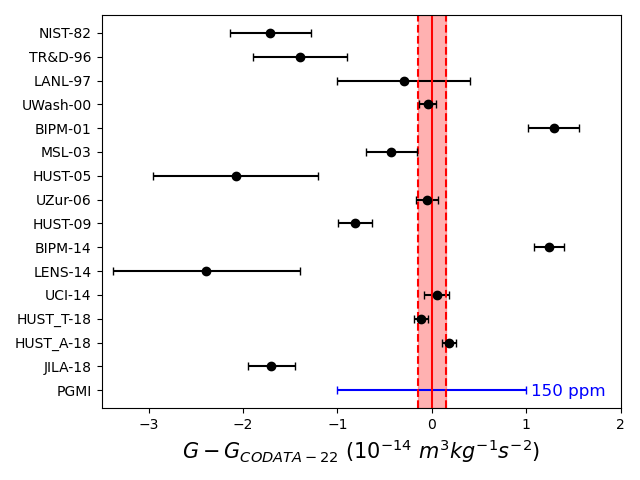}
    \caption{
    Difference between experimental results of G and the CODATA 2022 recommended value, $G=\SI{6.67430\pm0.00015e-11}{\meter\cubed\per\kilo\gram\per\second\squared}$.
    The shaded red region denotes the uncertainty in the CODATA 2022 value of G.
    The blue errorbar shows the uncertainty in G from the neutron PGMI experiment proposed in this work.
    Uncertainties are shown to 1 standard deviation.
    }
    \label{fig:codata2022-G-vals}
\end{figure}

Performing a precision measurement of the Newtonian constant of gravitation is a longstanding challenge, evident when comparing the relative uncertainty to which other fundamental constants of nature are known.
The recommended value for the G, $\SI{6.67430\pm0.00015e-11}{\meter\cubed\per\kilo\gram\per\second\squared}$, is listed with a relative uncertainty of $2.2\cdot 10^{-5}$, a significantly larger uncertainty than, for example, the Rydberg Constant, $R_\infty$, listed at $1.1\cdot 10^{-12}$, and the vacuum electric permittivity, $\epsilon_0$, at $1.6\cdot 10^{-10}$~\cite{mohr2024codata}.
Most measurements of G attempt to measure the gravitational force using a pendulum or force balance with source masses on the order of $\SI{10}{\kilo\gram}$.
Examples include the historical Cavendish pendulum~\cite{Boys1890I.Experiment}, the time-of-swing torsion pendulum~\cite{heyl1930redetermination}, and the more modern angular acceleration method ~\cite{Rose1969DeterminationG, Gundlach1996NewPresented, Gundlach1999AConstant, Gundlach2000MeasurementFeedback}.
Other experiments have used electrostatic and torsion force balances~\cite{Quinn2001AMethods, Quinn2013ImprovedMethods, Quinn2014TheG}, a Fabry-P\'{e}rot to measure changes in pendulum frequencies~\cite{Schurr1991ALaw, Kleinevo1999AbsoluteG}, and laser-cooled atom interferometry~\cite{Prevedelli2014MeasuringInterferometer, Rosi2014PrecisionAtoms} while a recent measurement utilized resonating beams to extract a dynamic measurement of G~\cite{brack2022dynamic}.
Although the reported uncertainty of modern measurements has been reduced~\cite{Li2018MeasurementsMethods}, significant discrepancies between measurements used to generate the CODATA estimate of G persist, as shown in Figure~\ref{fig:codata2022-G-vals}.~\cite{mohr2024codata, NSFNIST}.
In the two centuries since Cavendish's experiment, the relative uncertainty in the gravitational constant has only improved by two orders of magnitude, with disagreement between experiments of $550$ parts-per-million (ppm)~\cite{Xue2020PrecisionConstant}.
This necessitated an expansion factor of $3.9$ applied to the uncertainties from each experiment shown in Figure~\ref{fig:codata2022-G-vals} when calculating the 2022 CODATA recommended value for G~\cite{mohr2024codata}.
This underscores the need for experiments to measure G with systematic uncertainties that can be decoupled from existing measurements.

Modern metrology often makes use of matter-wave interferometers where sub-atomic particles~\cite{Rauch1974TestInterferometer}, atoms~\cite{Peters2001High-precisionInterferometry}, and even molecules~\cite{Brezger2002Matter-WaveMolecules} are split in a quantum superposition.
Neutrons are convenient probes of gravity as they are both massive quantum particles and electrically neutral~\cite{Werner2001NeutronMechanics, Cronin2009OpticsMolecules}.
Neutron interferometry has historically been used in several groundbreaking experiments,  such as the first observation of gravitational effects on a quantum particle~\cite{Colella1975ObservationInterference} and the first measurement of the $4\pi$ symmetry of spin-$\sfrac{1}{2}$ particles~\cite{Rauch1975VerificationFermions}.
Gravitational effects on neutrons were measured by Colella et al. in 1974 with a perfect-crystal neutron interferometer (NI)~\cite{Colella1975ObservationInterference}, and by van der Zouw et al. in 2000 using a phase grating NI~\cite{vanderZouw2000AharonovBohmInterferometer}.
However, single-crystal interferometers are limited to monoenergetic neutrons and the phase grating NI of~\cite{vanderZouw2000AharonovBohmInterferometer} requires very cold, low-flux neutrons.
Therefore both interferometers, restricted by low neutron flux, could not generate the statistics required for precision measurements of gravity.

It has been shown that a series of nano-fabricated phase gratings can be used to create a polychromatic far-field interferometer with x-rays~\cite{Miao2016AImaging} and neutrons~\cite{Sarenac2018ThreeApplications, Pushin2017Far-fieldImaging, Heacock2019AngularVisibility}, called a phase-grating moiré interferometer (PGMI).
Of interest here is a PGMI using three phase gratings, with phase heights of $\pi/2$, $\pi$, and $\pi/2$.
In a 3-PGMI, a fringe pattern is created by interfering the first and zeroth diffraction orders from the first grating, as described in Ref.~\cite{Miao2016AImaging} and shown in Figure~\ref{fig:pgmi-paths}.
The increased wavelength acceptance of the PGMI, coupled with interferometer areas comparable to single-crystal NIs, allows the PGMI to measure significantly weaker forces than previously possible.
A PGMI measurement of the gravitational force from a source mass whose density profile can be measured precisely provides a method of measuring the gravitational constant with systematic effects independent of measurements in the most recent CODATA report~\cite{mohr2024codata}.

We describe an experiment that can measure gravitational effects using the three-grating arrangement of the PGMI and model the expected phase shift induced by a source mass.
This model is validated by simulating existing neutron phase grating interferometric experiments~\cite{vanderZouw2000AharonovBohmInterferometer, Pushin2017Far-fieldImaging, Sarenac2018ThreeApplications}.
We then apply this model to quantify systematic effects and uncertainty from several significant sources, demonstrating the feasibility of measuring G using a neutron PGMI to a precision of $\SI{150}{ppm}$.

\section{Phase Grating Moir\'{e} Interferometer Model}
\label{sec:model}

In the model described here, we calculate the phase shift accumulated along each neutron's propagation axis from the first grating to the neutron camera.
The motivation to develop a new model describing the PGMI is the inability of other models to simulate non-linear potentials, like those produced by the gravitation of a source mass, efficiently.
Several neutron PGMI configurations have been previously simulated using a momentum space model of each grating and propagation effects~\cite{sarenac2023cone}.
Here, we develop a new model of the neutron PGMI that can efficiently compute the effect of neutron propagation through an arbitrary potential.

\begin{figure*}
    \centering
    \includegraphics[width=\linewidth]{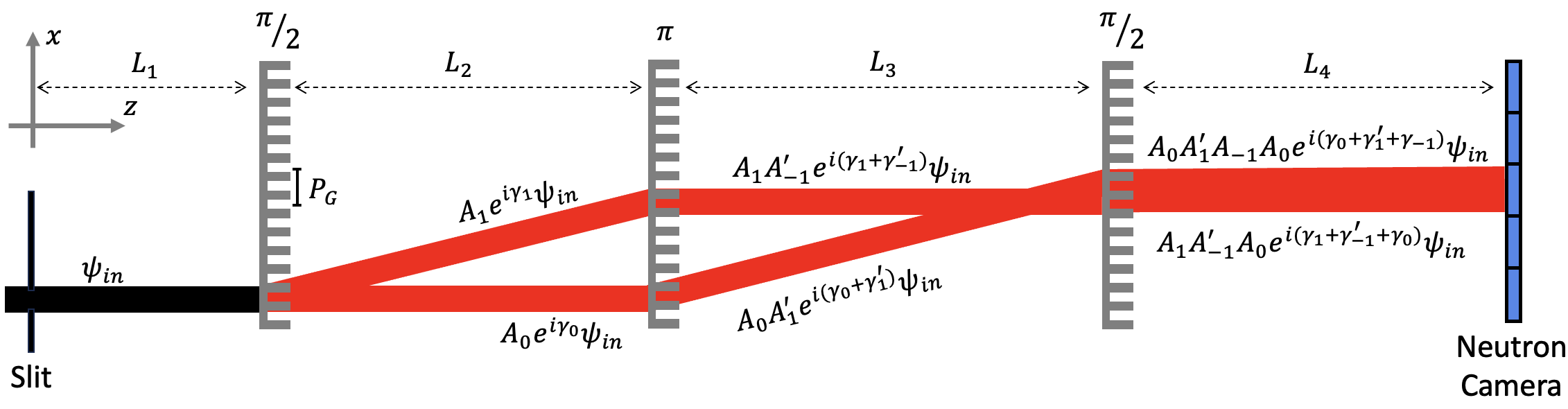}
    \caption{
    Schematic showing interfering paths in a three-grating neutron PGMI as red lines.
    The source neutron slit is shown in black on the left while the blue rectangle on the right indicates the camera position.
    The wavefunctions of paths within the PGMI are shown with each path's amplitude ($A_l$) and phase ($\xi_l$) factors.
    Amplitude and phase factors with a prime ($A_l'$, $\xi_l'$) denote values from the transmission function of the $\pi$-height grating.
    The terms shown beside each path segment corresponds to the wavefunction at the next object in the beam (grating/camera) and thus the phase term ($\gamma_i$) includes the phase accumulated from propagation.
    }
    \label{fig:pgmi-paths}
\end{figure*}

A phase grating separates a neutron wave-packet into several diffraction orders, which may be further separated by subsequent gratings.
The probability of detecting a neutron at a particular pixel is given by the amplitude of each path and the phase accumulated along the trajectories that reach said pixel.
For pixels with a single trajectory, the probability of detecting a neutron is given by the product of each diffraction order amplitude, $I_p = \left| A_l \cdot A'_m \cdot A_n \right|^2$ where the amplitude, $A_n$ denotes the $n$-th diffraction order from a grating and the $\pi$-height grating is distinguished from the $\pi/2$-height by the prime, $A'_m$.
An example case of multiple trajectories that reach a given pixel is illustrated in Figure~\ref{fig:pgmi-paths}.
The probability of a neutron being detected by a pixel at position $x_p$, with multiple neutron trajectories reaching said pixel, is

\begin{equation}
    I(x_p) = \left|\sum_{\{l,m,n\}} A_l A'_m A_n e^{-i (\gamma_l + \gamma'_m + \gamma_n)} \right|^2,
\label{eqn:path-amplitude}
\end{equation}

\noindent
where the terms $\gamma_i$ include the phase accumulated by propagation.
For convenience, we combine these amplitudes and phases into $A_{lmn} \equiv A_l \cdot A'_m \cdot A_n$, and $\gamma_{lmn} \equiv \gamma_l + \gamma'_m + \gamma_n$.
The phase terms in Eqn~\ref{eqn:path-amplitude} can be computed following the Wentzel–Kramers–Brillouin (WKB) approximation, commonly applied to neutron interferometers~\cite{Werner2001NeutronMechanics, ekstrom1995measurement, perreault2006measurement}.
Therefore, the phase $\gamma_{lmn}$ can be computed using a path-integral along the neutron trajectory $S_{lmn}$

\begin{equation}
    \gamma_{lmn} = \int_{S_{lmn}} \mathbf{k(r)} \cdot d\mathbf{r},
\label{eqn:path-integral}
\end{equation}

\noindent
where the neutron wave-vector $\mathbf{k(r)}$ depends on the potential at any point in space.
A Monte Carlo simulation of the 3-PGMI, computing the phases along all possible paths, should reproduce the interference pattern at the camera, where the neutron angle, wavelength, and position at the source slit are varied.
Our model has been validated against existing phase grating interferometer experiments, with gravitational potentials and in PGMI arrangements, in Appendix~\ref{sec:model-validation}.

\section{Measuring G}

The Hamiltonian for a neutron propagating between the gratings of a 3-PGMI is

\begin{equation}
    \mathcal{H} = \frac{\hbar^2 \mathbf{k}^2}{2 m_N} - U_G(\mathbf{r}) - U_S(\mathbf{r}) - U_A(\mathbf{r}) - U_L(\mathbf{r}, t),
    \label{eqn:test-mass-ham}
\end{equation}

\noindent
where $\hbar$ is the reduced Plank's constant, $m_N$ is the neutron mass, $\mathbf{r}$ is the neutron position coordinate, $U_G$ is the gravitational potential of the source mass, $U_S$ is the correction to the Hamiltonian due to Earth's rotation, $U_A$ is the nuclear potential of the ambient air, and $U_L$ is the potential induced by lunar tidal forces, and $t$ is the time coordinate.
The phase accumulated from the source slit to the camera can then be computed with Eqn.~\ref{eqn:path-integral} along the each neutron's propagation axis, with the neutron wavevector

\begin{align}
    \mathbf{k} &= \mathbf{k}_0 \sqrt{1 + \frac{2m_NU_\textnormal{tot}}{\hbar^2 \mathbf{k}_0^2}} \nonumber\\
    &\approx \mathbf{k}_0 \left(1 + \frac{m_NU_\textnormal{tot}}{\hbar^2 \mathbf{k}_0^2} + \mathcal{O} \left[ \left( \frac{U_\textnormal{tot}}{E_0} \right)^2 \right]  \right),
\label{eqn:wave-vector}
\end{align}

\noindent
where the potentials in Eqn.~\ref{eqn:test-mass-ham} have been collected into $U_\textnormal{tot}$, and $E_0$ is the initial kinetic energy of the neutron.
As $U_\textnormal{tot} << E_0$, the accumulated neutron phase with is

\begin{equation}
    \gamma_{lmn} \approx \int_{S_{lmn}} \mathbf{k}_0 \cdot d\mathbf{s} + \frac{m_N}{\hbar^2 \mathbf{k}_0^2}\int_{S_{lmn}} U_\textnormal{tot}(s)~\mathbf{k}_0 \cdot d\mathbf{s},
\label{eqn:path-integral-potential}
\end{equation}

\noindent
which implies that, to the first order, the phase induced by each of the potentials in Eqn.~\ref{eqn:test-mass-ham} can be written in separate terms, as is done for models of precision atom interferometry experiments~\cite{holmgren2010absolute}.
As described in the Appendix of~\cite{Miao2016AImaging}, the interference pattern generated by the PGMI arises from two paths that are detected at the same pixel of the detector regardless of the neutron's incident angle or energy.
For two interfering paths, the measured intensity given by Eqn.~\ref{eqn:path-amplitude} reduces to

\begin{align}
    I_G(x_p) \sim &\cos\left( \Delta\gamma_0 + \Delta\gamma_G + \Delta\gamma_S + \Delta\gamma_A + \Delta\gamma_L \right) \\
    &\equiv \cos\left[ \Delta\Phi_G(x_p) \right] \nonumber,
\end{align}

\noindent
where the phase differences $\Delta\gamma_i$ correspond to the potential fields in Eqn.~\ref{eqn:test-mass-ham}.
A measurement of the gravitational constant will require isolating the phase difference induced by the gravitational potential of the source mass, $\Delta\gamma_G$, from the other phase factors. 
This can be achieved by subtracting the measured phase from a null experiment, $\Delta\Phi_N(x_p)$, where the source mass is removed such that its gravitational potential is negligible.
The phase shift induced by the gravitational potential can then be extracted:

\begin{align}
    \Delta\gamma_G &= (\Delta\Phi_G - \Delta\gamma_S - \Delta\gamma_{A, G} - \Delta\gamma_{L, G}) \nonumber\\
    &\hspace{1cm}- (\Delta\Phi_{N} - \Delta\gamma_S - \Delta\gamma_{A, N} - \Delta\gamma_{L, N})\nonumber\\
    &\equiv (\Delta\Phi_G - \Delta\gamma_{\textnormal{corr}, G}) - (\Delta\Phi_{N} - \Delta\gamma_{\textnormal{corr}, N}),
\label{eqn:gravity-phase}
\end{align}

\noindent
where the subscripts $G$ and $N$ on each phase difference $\Delta\gamma_{i, G/N}$ refer to the source mass and null experiment, respectively. 
Note that different correction factors are required as ambient atmospheric conditions and tidal forces will vary between null and source mass experiments.
The phase difference at the camera coordinate $x_p$ due to the source mass can be written from Eqn.~\ref{eqn:path-integral-potential},

\begin{equation}
    \Delta\gamma_G(x_p) = \frac{m_N}{\hbar^2 \mathbf{k}_0^2}\oint_{S_{x_p}} G \cdot m_N \left[\iiint \frac{\rho(\mathbf{r}_V)}{\mathbf{s}-\mathbf{r}_V} d^3 \mathbf{r}_V\right] ~\mathbf{k}_0 \cdot d\mathbf{s},
\label{eqn:test-mass-phase}
\end{equation}

\noindent
where the path integral is over the specific path $S_{x_p}$ that reaches the camera at coordinate $x_p$, the volume integral is over the source mass-centered coordinate $\mathbf{r}_V$, and the density of the source mass material is described by $\rho(\mathbf{r}_V)$.
The gravitational potential term in Eqn.~\ref{eqn:test-mass-phase} can be solved numerically but has been solved analytically for right rectangular prisms, shown in Appendix~\ref{sec:prism-potential}~\cite{nagy2000gravitational, karcol2019two}.
Combining Equations~\ref{eqn:gravity-phase} and ~\ref{eqn:test-mass-phase}, the gravitational constant can be measured by computing

\begin{equation}
    G = \frac{\hbar^2 \mathbf{k}_0^2}{m_N^2} \frac{(\Delta\Phi_G - \Delta\gamma_{\textnormal{corr}, G}) - (\Delta\Phi_N - \Delta\gamma_{\textnormal{corr}, N})}{\oint_{S_{x_p}}\left[\iiint \frac{\rho(\mathbf{r}_V)}{\mathbf{s}-\mathbf{r}_V} d^3 \mathbf{r}_V\right] ~\mathbf{k}_0 \cdot d\mathbf{s}},
\label{eqn:big-g-result}
\end{equation}

\noindent
where the phase functions $\Delta\Phi_{G/N}$ are measured from the interference patterns of the source mass/null experiments, and the path integral term is computed for each coordinate across the camera.
This result applies to one particular pair of interfering paths (i.e., the $\{0, 1, -1\}$ and $\{1, -1, 0\}$ paths shown in Figure~\ref{fig:pgmi-paths}) at one neutron wavelength. 

In reality, a neutron 3-PGMI produces more than two pairs of interfering paths at each pixel of the detector, and accepts a distribution of neutron wavelengths.
Therefore, computing a value for G, as in Eqn.~\ref{eqn:big-g-result}, requires calculating the averaged phase from each potential.
A similar procedure is used in atom interferometry for measuring the electric polarizability using atom beams with a velocity distribution~\cite{ekstrom1995measurement, miffre2006measurement, holmgren2010absolute}.
For a polychromatic neutron 3-PGMI, the intensity is the incoherent sum over pairs of interfering paths and the wavelength distribution of the beam.
Rewriting Eqn.~\ref{eqn:path-amplitude}:

\begin{align}
    I(x_p) = \sum_{\substack{\{lmn, \\\tilde{l}\tilde{m}\tilde{n}\}}} & \int_0^\infty P(\lambda) \Big| A_{lmn} e^{-i\gamma_{lmn}} \nonumber\\
    &\hspace{1cm}+ A_{\tilde{l}\tilde{m}\tilde{n}} e^{-i\gamma_{\tilde{l}\tilde{m}\tilde{n}}} \Big|^2 d\lambda,
\label{eqn:wavefunction-averaging}
\end{align}

\noindent
where $\{lmn, ~\tilde{l}\tilde{m}\tilde{n}\}$ denotes a pair of interfering paths, and $P(\lambda)$ is the probability of a neutron having wavelength $\lambda$. 
As interference fringes arise exclusively from pairs of interfering paths of the 3-PGMI, we can simplify the modulus term to

\begin{align}
    I(x_p) &= \sum_{\substack{\{lmn, \\\tilde{l}\tilde{m}\tilde{n}\}}} \left(|A_{lmn}|^2 + |A_{\tilde{l}\tilde{m}\tilde{n}}|^2\right) \int_0^\infty P(\lambda) \big\{ 1 \nonumber\\
    &\hspace{2.5cm} + \mathcal{C}(\lambda) \cos\left[ \Delta\gamma(\lambda) \right] \big\} d\lambda \nonumber\\
    &\equiv I_0(x_p) \left\{1 + \langle\mathcal{C}\rangle \cos\left[ \left\langle \Delta\Phi_{G/N}(x_p) \right\rangle \right] \right\},
\label{eqn:cosine-averaging}
\end{align}

\noindent
where the sum is over all interfering path pairs, $\Delta\gamma$ and $\mathcal{C}$ are the wavelength-dependent phase and contrast, $I_0(x_p)$ is the average intensity profile across the camera, and $\langle~\rangle$ denote the average values. 
From this, we see a different measured phase, $\left\langle \Delta\Phi_{G/N}(x_p) \right\rangle$, for the null and source mass measurements.
Note that these averages are not simple averages over the wavelength distribution and interfering pairs but result from solving the integral in Eqn.~\ref{eqn:cosine-averaging} numerically~\cite{miffre2006measurement}.
These averages can be similarly applied to find the resulting phase with Equation~\ref{eqn:big-g-result}, replacing the phase values $\Delta\Phi$, corrections $\Delta\gamma$, and path integral terms with their averaged equivalents.
Computing G from this experiment,

\begin{equation}
    G = \frac{\hbar^2}{m_N^2} \frac{(\Delta\Phi_G - \langle\Delta\gamma_{\textnormal{corr}, G}\rangle) - (\Delta\Phi_N - \langle\Delta\gamma_{\textnormal{corr}, N}\rangle)}{\left\langle \mathbf{k}_0^{-2}\oint_{S_{x_p}}\left[\iiint \frac{\rho(\mathbf{r}_V)}{\mathbf{s}-\mathbf{r}_V} d^3 \mathbf{r}_V\right] ~\mathbf{k}_0 \cdot d\mathbf{s} \right\rangle},
\label{eqn:big-g-result-avg}
\end{equation}

where the average in the denominator is over all interfering paths that reach a particular pixel and over the wavelength distribution of the neutron beam.

\section{Proposed Experiment}
\label{sec:proposal}

\begin{figure*}
    \centering
    \includegraphics[width=\linewidth]{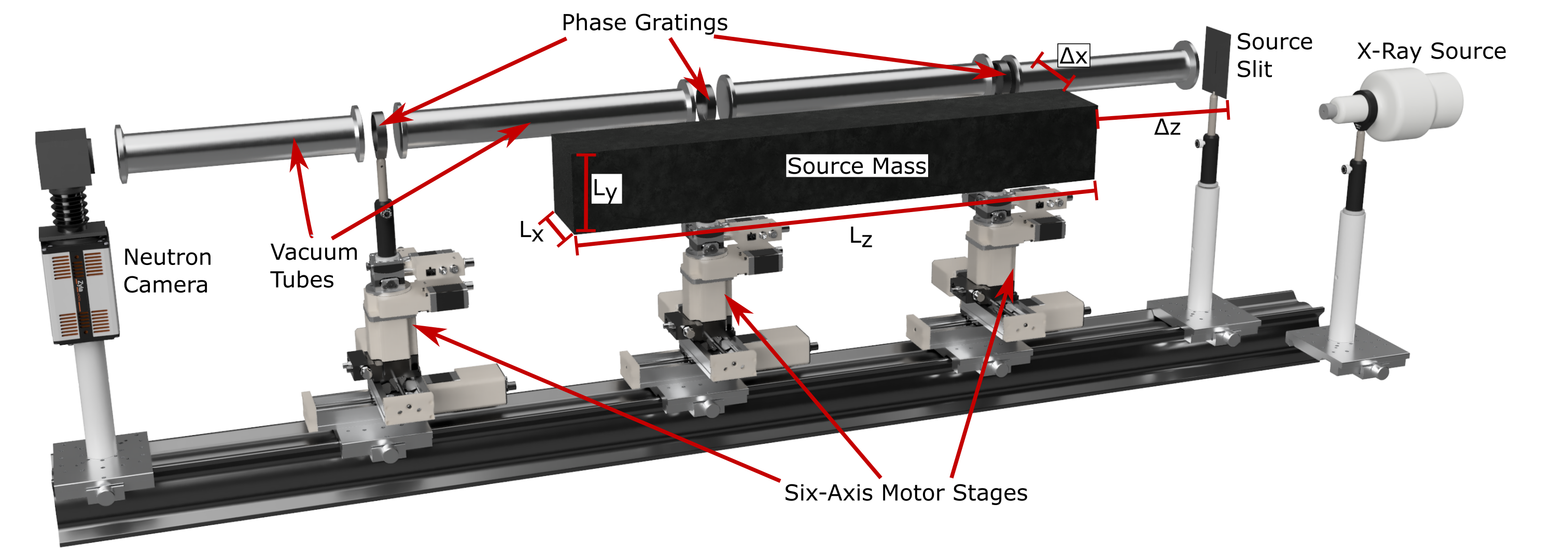}
    \caption{
    Model of the apparatus proposed in this work to measure G using a neutron PGMI.
    This design features an x-ray source that can be used when the neutron beam is blocked, allowing for alignment validation independent of gravitational effects.
    The design of our neutron camera, similar to that shown by LaManna et al.~\cite{lamanna2017neutron}, can image both neutrons and x-rays.
    }
\label{fig:apparatus}
\end{figure*}

Figure~\ref{fig:apparatus} shows the proposed arrangement of three phase gratings, each mounted on a six-axis motor control stage.
The use of vacuum tubes between the gratings is motivated in Section~\ref{sec:atmosphere}.
This design allows an x-ray source to be used when the neutron beam is blocked to monitor changes in grating alignment between null and source mass experiments independent of gravitational effects.
As our neutron camera can image both neutrons and x-rays~\cite{lamanna2017neutron}, this x-ray source can be used to monitor changes in grating alignment without disturbing the phase gratings or camera.
The NG-6U neutron beam line at NCNR, where proof-of-concept experiments will take place, can accommodate a total interferometer length of up to $\SI{10}{\meter}$.
A measurement of G with this apparatus will take advantage of a polychromatic beamline, where we can achieve a neutron fluence rate of $\SI{e9}{ \per\cm\squared\per\second}$~\cite{hussey2015new}.
To achieve appropriate coherence over one cycle of $\SI{1}{\um}$ period gratings, patterned over a large area, we can use a neutron source slit of up to $\SI{35}{\mm} \times \SI{0.8}{\mm}$.

A $\SI{1}{\tonne}$ source mass was chosen for this design to maximize the gravitational potential across the 3-PGMI without requiring specialized equipment to move the mass.
The choice of source mass material should be motivated by two factors: high density and precisely measurable density variations.
The transparency of PbWO$_4$ ($\rho = \SI{8235}{\kilo\gram\per\meter\cubed}$) has allowed for optical characterizations of the material to a relative density uncertainty of $\SI{1.16}{\micro\gram\per\centi\meter\cubed}$~\cite{assumin2022imaging}.
Solid lead ($\rho = \SI{11342}{\kilo\gram\per\meter\cubed}$) will create a larger phase shift due to its higher density, albeit with a greater uncertainty in source mass density. 
Neutron phase contrast imaging has been used to measure the density distribution of PbWO$_4$, achieving a bound on the density gradient of $\frac{1}{\rho_N}\frac{d\rho_N}{dx}<\SI{0.5e-6}{\per\centi\meter}$, where $\rho_N$ is the number density of the material~\cite{assumin2022neutron}.
This uncertainty in the density gradient is $25$ times greater than what was achieved with optical characterization, corresponding to an uncertainty in density of $\SI{100}{\micro\gram\per\centi\meter\cubed}$ across the source mass width.
With these density uncertainty values, the model outlined in Section~\ref{sec:model} shows that a PbWO$_4$ source mass results in a relative uncertainty in G of $\SI{0.13}{ppm}$ while a solid lead source mass produces a relative uncertainty of $\SI{2}{ppm}$.
The increased uncertainty in lead density is a reasonable compromise given the increased gravitational phase shift compared to lead tungstate.
It should be noted that these experiments measuring density using optical and neutron methods used samples $\SI{2.3}{\centi\meter}\times\SI{2.3}{\centi\meter}\times\SI{12}{\centi\meter}$ in size.
Applying these techniques may require a source mass composed of smaller masses arranged to form a larger rectangle.
This will likely lead to additional sources of uncertainty based on temperature variation and alignment, dependent on the exact size and number of masses used.

The location and dimensions of the source mass significantly affects the expected phase shift $\Delta\gamma_G$, and therefore affects the expected beam time for a desired measurement precision.
We can apply the path integral model to compute the expected phase shift and contrast from a particular arrangement of the source mass and phase gratings, allowing for a minimization of the expected beam time.
The relationship between phase uncertainty, interference fringe contrast, and beam time is described in Appendix~\ref{sec:fit-stats}, demonstrating that minimizing beam time is equivalent to maximizing the product of fringe contrast and target phase uncertainty $\mathcal{C}\cdot\sigma_{\Delta\Phi}$.
The displacements $\Delta x$, $\Delta y$, and $\Delta z$ (listed in Table~\ref{tab:parameter-uncertainty}) are measured from the surface of the source mass to the center of the source slit. 
The displacements $\Delta x$ and $\Delta z$ are measured to the nearest face of the source mass, while $\Delta y$ is measured from the top surface to the center of the beam.
The result of this optimization is that a source mass of dimensions $\{L_x, L_y, L_z\} = \{\SI{13.0}{\cm}, \SI{13.0}{\cm}, \SI{500}{\cm}\}$ should be positioned with $\Delta z = 0$ (see Fig.~\ref{fig:apparatus}) beside a grating interferometer with lengths $\{L_1, L_2, L_3, L_4 \} = \{\SI{80.0}{\centi\meter}, \SI{200.0}{\centi\meter}, \SI{200.9}{\centi\meter}, \SI{399.1}{\centi\meter}\}$.

\begin{figure}
    \centering
    \includegraphics[width=\linewidth]{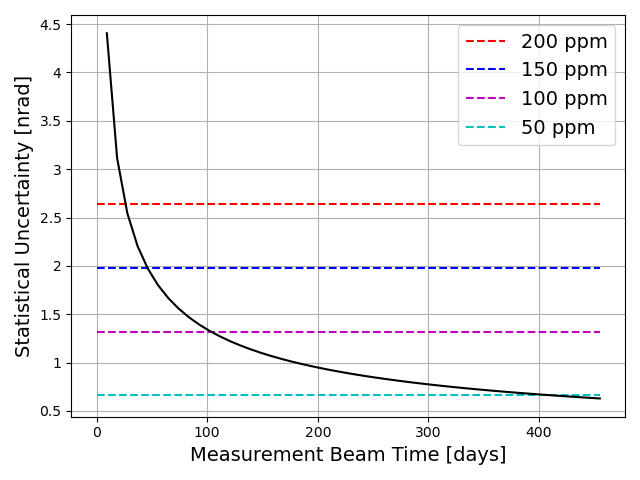}
    \caption{
    Statistical uncertainty in the phase of an interference pattern created in a neutron 3-PGMI with the beam parameters in Table~\ref{tab:parameter-uncertainty}.
    Dotted lines show the $\SI{50}{ppm}$, $\SI{100}{ppm}$, $\SI{150}{ppm}$, and $\SI{200}{ppm}$ relative uncertainties in the measured phase $\Delta\Phi_{G/N}$.
    }
    \label{fig:stat-uncert}
\end{figure}

The optimized interferometer parameters result in a contrast of $\mathcal{C} = \SI{22.1}{\percent}$, both with and without the source mass, and phase shift due to gravity of $\phi = \SI{13.2}{\micro\radian}$.
While the 3-PGMI is capable of producing higher interference fringe contrast, the arrangement described here produced the maximum product of $\mathcal{C}\cdot\sigma_{\Delta\Phi}$.
The statistical uncertainty as a function of the beam time (for each of the null/source mass experiments) is shown in Figure~\ref{fig:stat-uncert}, with relative uncertainties of $\SI{50}{ppm}$, $\SI{100}{ppm}$, $\SI{150}{ppm}$, and $\SI{200}{ppm}$ denoted with dotted lines.
A measurement of this phase to $\SI{92.7}{ppm}$ results in an expected beam time of $\SI{120}{\day}$ and a combined statistical uncertainty for both the null and source mass measurements of $\SI{131.1}{ppm}$.

\section{Uncertainties}
\label{sec:uncertainity}

The uncertainty in G from Equation~\ref{eqn:big-g-result-avg}, depends on the uncertainties in the phase correction terms $\langle \Delta \gamma_{\textnormal{corr}, G/N} \rangle$, the source mass-induced phase $\langle \Delta\gamma_G \rangle$, and the measured phase profiles $\Delta\Phi_{G/N}$.
The model described here allows us to quantify these uncertainties from uncertainties in the underlying experimental parameters.
The uncertainty in each of these parameters is summarized in Table~\ref{tab:parameter-uncertainty}.
The total uncertainty in G and the relative uncertainty due to each effect are summarized in Table~\ref{tab:uncertainty-budget}.

\subsection{Phase Correction Uncertainty}
\subsubsection{Atmospheric Phase}
\label{sec:atmosphere}

As outlined in chapter 3.4.1 of ~\cite{Werner2001NeutronMechanics}, atmospheric corrections depend on pressure, temperature, and humidity.
The phase shift due to humidity and air density is

\begin{equation}
    \Delta\gamma_A = -\lambda \Delta S \frac{\mathcal{L} P}{R T} \frac{\sum_i N_i b_{c,i}}{\sum_i N_i},
\end{equation}

\noindent
where the sums are over each isotope $i$ in the ambient atmosphere, $\Delta S$ is the path length difference, $N_i$ and $b_{c, i}$ are the number density and scattering length for a particular isotope, $\mathcal{L}$ is the Loschmidt number, $R$ is the ideal gas constant, $P$ is the ambient pressure, and $T$ is the ambient temperature.
The difference in phase between vacuum and 1 atmosphere, $\SI{20}{\degreeCelsius}$, and $43\%$ humidity air for the PGMI with $\SI{4}{\angstrom}$ neutrons is $\Delta\gamma_A = \SI{3.9}{\micro\radian}$.
Monitoring these parameters to $1\%$ uncertainty results in a relative uncertainty in G of $\SI{129.8}{ppm}$.
Placing evacuated cylinders, each at a pressure of $\SI{1}{\pascal}$, between the source slit, gratings, and camera will reduce the phase due to atmospheric effects to $\SI{3.8E-11}{\radian}$ which, without any correction between the null and source mass experiments, results in a relative uncertainty in G of $\SI{2.72}{ppm}$.

\subsubsection{Sagnac Phase}

Non-inertial effects on neutrons have been well-documented, and the effect of the Earth's rotating reference frame can be computed from elementary mechanics~\cite{staudenmann1980gravity, page1975effect, werner1979effect, atwood1984neutron}. 
The phase accumulated by a neutron as the Earth rotates about a vector $\mathbf{\Omega}$ within an interferometer with an area vector $\mathbf{A}$ is

\begin{equation}
    \Delta\gamma_S = \frac{2 m_N}{\hbar} \mathbf{\Omega} \cdot \mathbf{A}.
\end{equation}

\noindent
If we assume that the interferometer is pointing perpendicular to the surface of the Earth, this dot product depends only on the latitude angle $\theta_L$. The area of a 3-PGMI is dependent on the angle between the first and zeroth diffraction orders of the first grating, therefore the Sagnac phase depends on wavelength as 

\begin{equation}
    \Delta\gamma_S = \frac{2 m_N}{\hbar} \big| \mathbf{\Omega} \big| \left( \frac{L}{2} \right)^2 \left( \frac{\lambda}{P_g} \right) \cos \left( \theta_L \right),
    \label{eqn:sagnac_phase}
\end{equation}

\noindent
where $L$ is the total interferometer length, and $P_g$ is the grating period.
While the Sagnac effect results in a $\SI{19.9}{\radian}$ phase shift, this is present in both the null and source mass measurements.
Therefore, the uncertainty in G due to the Sagnac effect is $\SI{10}{ppm}$, arising from differences in the effective wavelength distribution between measurements.

\subsubsection{Lunar Gravitational Phase}

The phase induced by tidal forces can be computed by treating the Moon as a point mass and calculating the distance from the laboratory reference frame to the lunar center of mass.
Note that these tidal forces are due to changes in the position of the Moon and not changes in local gravitational acceleration caused by sea level variation.
The Earth-Moon distance has been measured to millimeter precision by the Apache Point Observatory Lunar Laser-ranging Operation~\cite{battat2009apache}, along with measurements of Earth's obliquity to the ecliptic and the lunar orbital inclination at relative precisions $< \SI{1}{ppm}$~\cite{luzum2011iau, park2021jpl}.
The remaining parameters used to compute the lunar gravitational phase are the latitude and orientation of the PGMI apparatus, the rotation phase of the Earth, and the orbital phase of the Moon.
These parameters allow us to calculate the lunar phase shift by integrating the lunar gravitational potential along each neutron trajectory, as described by Eqn.~\ref{eqn:path-integral-potential}.
Using commercial GPS and electronic-compass devices, the latitude and orientation relative to Earth's rotation axis can be measured with an uncertainty of $\ang{;;0.14}$ and $\ang{0.5}$, respectively~\cite{van2015world, jewell2024ecs, NIST_disclaimer}.
Combining these sources of uncertainty into a Monte Carlo simulation, the relative uncertainty in G from lunar gravitation is $\SI{1.7}{ppm}$.
The additional effects caused by tidal changes in sea level can be controlled for with concurrent gravimeter measurements and tidal modeling.

Precision measurements of local acceleration using atom interferometry have shown that gravity tides produce measurable variations in $g$ on the order of $\SI{1}{ppm}$~\cite{Peters2001High-precisionInterferometry}.
Despite this, only two of the sixteen experiments contributing to the CODATA 2022 recommended value of Newton's Constant of Gravitation address lunar gravitation~\cite{mohr2024codata}.
Both ~\cite{schlamminger2006measurement} and ~\cite{Prevedelli2014MeasuringInterferometer} refer to ``tidal effects'' or ``tidal forces'', though no value on the uncertainty from these effects is given.
To estimate the effect of tidal forces on a torsion pendulum, we can apply our lunar potential simulation to the arrangement in Luther and Towler~\cite{luther1982redetermination}.
For a torsion balance measuring the gravitational force from two source masses, G is calculated with

\begin{equation}
    G = \cfrac{\Delta(\omega^2) I}{k_g},
\end{equation}

\noindent
where the torsion pendulum has a momentum of inertia $I$, twists about an angle $\theta$, and the gravity of the source mass produces a restoring force with a reduced torsion constant $k_g$.
That fictitious torsion constant is calculated by $k_g = G^{-1} d^2 U_T/d\theta^2$, where the source mass potential $U_T = G M m / R$ is integrated over the test-mass volume.
The effect of tidal forces can be calculated by adding the potential due to the Moon's gravitation to the term $U_T$.
Using the listed uncertainty in test-mass density, there is a significant torque induced by lunar gravitation, which produces an additional uncertainty in G of $\SI{39.5}{ppm}$ to the result in Ref.~\cite{luther1982redetermination}.

\subsection{Source-Mass Phase Uncertainty}

The uncertainty of each experimental parameter is shown in Table~\ref{tab:parameter-uncertainty}.
Each of these parameters is varied in a Monte Carlo simulation, and the resulting uncertainty in G is examined here.

\begin{table*}
    \centering
    \begin{tabular}{l|c|c|r}
        Parameter & Symbol & Value & Uncertainty \\
        \hline \hline
        Source-Mass Dimension & $L_x$, $L_y$, $L_z$ & \SI{13.0}{\cm}, \SI{13.0}{\cm}, \SI{500}{\cm} & \SI{5}{\micro\meter} \\
        Mass Density & $\rho$ & \SI{11342}{\kilo\gram\per\meter^3} & $\SI{1.16}{\micro\gram\per\centi\meter\cubed}$ \\
        Mass Surface Variation & $\delta L_x$, $\delta L_y$, $\delta L_z$ & 0 & \SI{5}{\micro\meter} \\
        \hline
        Ambient Temperature & T & \SI{20}{\degreeCelsius} & \SI{0.1}{\degreeCelsius} \\
        Ambient Pressure & $P$ & \SI{101}{\kilo\pascal} & \SI{1}{\kilo\pascal} \\
        Ambient Humidity & $\epsilon$ & \SI{43}{\percent} & \SI{0.5}{\percent} \\
        \hline
        Mass Displacement (Horizontal) & $\Delta x$ & \SI{20}{\centi\meter} & \SI{3}{\micro\meter} \\
        Mass Displacement (Vertical) & $\Delta y$ & \SI{6.5}{\cm} & \SI{0.5}{\milli\meter} \\
        Mass Displacement (Beam-axis) & $\Delta z$ & \SI{80.0}{\centi\meter} & \SI{0.5}{\milli\meter} \\
        Grating Separations & $L_1$, $L_2$, $L_3$, $L_4$ & \SI{80.0}{\centi\meter}, \SI{200.0}{\centi\meter}, \SI{200.9}{\centi\meter}, \SI{399.1}{\centi\meter} & \SI{3}{\micro\meter} \\
        \hline
        Mean Grating Period & $P_g$ & $\SI{700}{\nano\meter}$ & $\SI{41}{\pico\meter}$ \\
        Mean Neutron Wavelength & $\Bar{\lambda}$ & $\SI{4}{\angstrom}$ & $\SI{0.0004}{\angstrom}$ \\
        \hline
        Neutron Flux & $\dot{N}$ & \SI{e9}{\per\cm\squared\per\second} & \SI{3e4}{\per\cm\squared\per\second} \\
        Slit Size & $s_w$ & $\SI{35}{\mm} \times \SI{0.8}{\mm}$ & \SI{1}{\micro\meter} \\
    \end{tabular}
    \caption{
    Parameters for the experiment proposed in Sec.~\ref{sec:proposal} with absolute uncertainties to $1\sigma$ ($68\%$ confidence).
    The distances to each grating from the camera and source slit are each measured with the same absolute uncertainty as the grating separations.
    Source mass displacements ($\Delta x$, $\Delta y$, $\Delta z$) are given to the surface of the source mass.
    Parameters are grouped by relevance to subsections in Sec.~\ref{sec:uncertainity}.
    }
    \label{tab:parameter-uncertainty}
\end{table*}

\begin{table*}
    \centering
    \begin{tabular}{l|r}
        Parameter & Uncertainty in G (ppm) \\
        \hline \hline
        Grating Mean Period & 58 \\
        Neutron Wavelength & 40 \\
        Source-Mass Displacement (Horizontal) & 11 \\
        Source-Mass Displacement (Beam-axis) & 11 \\
        Sagnac Phase & 10 \\
        Grating Thermal Variation & 5.2 \\
        Source-Mass Displacement (Vertical) & 3.5 \\
        Atmospheric Effects & 2.7 \\
        Source-Mass Density & 2.0 \\
        Lunar Gravitation & 1.7 \\
        Grating Separations & 0.78 \\
        Source-Mass Surface Flatness & 0.53 \\
        \hline
        Systematic Uncertainty & 73  \\
        Statistical Uncertainty ($\SI{240}{\day}$) & 130  \\
        \hline \hline
        Total & 150
    \end{tabular}
    \caption{Uncertainty budget for the 3-PGMI measurement proposed using a rectangular source mass.
    The methods used to calculate the uncertainty of each parameter are described in Sec.~\ref{sec:uncertainity}, with values listed in this table to two significant digits.
    Statistical uncertainty of $\SI{131.1}{ppm}$ is the result of $\SI{240}{\day}$ of total experiment time consisting of two $\SI{120}{\day}$ beam time runs for a null and source mass experiment.
    }
    \label{tab:uncertainty-budget}
\end{table*}

\subsubsection{Displacements}

With an absolute uncertainty of $\SI{0.5}{\milli\meter}$ in beam-axis (vertical) displacements, the uncertainty in G is $\SI{11}{ppm}$ ($\SI{3.5}{ppm}$).
By lowering the gratings below the neutron beam height, the source mass can be translated into the beam, providing an accurate displacement zero in the horizontal direction.
Displacement sensors, like those offered by Attocube~\cite{NIST_disclaimer}, can measure the horizontal displacement to an uncertainty of $\SI{3}{\micro\meter}$, resulting in an uncertainty in G of $\SI{11}{ppm}$.
The same sensors can be used to measure the distances from the source slit to the first grating, between each grating, and from the third grating to the neutron camera.
Uncertainties of $\SI{3}{\micro\meter}$ in each of the distances $L_1$, $L_2$, $L_3$, and $L_4$, results in a $\SI{0.78}{ppm}$ relative uncertainty in G.

\subsubsection{Phase Grating Period}

Using small-angle x-ray scattering (SAXS)~\cite{hu2004small}, or by imaging the diffracted orders from an x-ray pencil beam, the mean grating period can be measured point-by-point across the gratings to $\SI{58}{ppm}$.
This uncertainty in the grating period, using the Monte Carlo simulation, leads to an uncertainty in G of $\SI{57.7}{ppm}$.
Thermal fluctuations during the experiment will cause expansion/contraction of the silicon wafers and therefore change the grating period. 
Assuming fluctuations of $\pm\SI{2}{\degreeCelsius}$, the uncertainty in G introduced by this change in grating period is $\SI{5.2}{ppm}$.

\subsubsection{Neutron Wavelength}

The uncertainty in the effective mean wavelength of a polychromatic beam can be calculated by a Monte Carlo error propagation of uncertainties in the wavelength and neutron count of a neutron spectroscopy experiment.
The error in the neutron count of a spectrometer follows Poisson statistics, while the uncertainty in the wavelength measured by a chopper experiment has been analyzed previously~\cite{vergara2023evaluation}.
Expanding the chopper spectrometer designed by Ref.~\cite{vergara2023evaluation} to $\SI{8}{\meter}$ length and using the parameters for the long spectrometer in that work, the weighted-averaged wavelength can be measured to \SI{98.4}{ppm}.
This results in an uncertainty in the source-mass-induced phase of 40.1 ppm.

\subsection{Statistical Uncertainty}
\label{sec:stats}

The final source of uncertainty in determining G from Eqn.~\ref{eqn:big-g-result-avg} is in the measured phase values, $\Phi_{G/N}$.
Assuming Poissonian distribution for the counts seen at each pixel of the detector, the uncertainty in the resulting least squares fit is

\begin{equation}
    (\sigma_{\Phi})^2 = \frac{1}{\dot{N} T\left(1 - \sqrt{1-\mathcal{C}^2} \right)},
\end{equation}

\noindent
where $\dot{N}$ is the count rate at the detector and $T$ is the measurement beam time.
A derivation of this result is shown in the Appendix~\ref{sec:fit-stats}.
The parameters in Table~\ref{tab:parameter-uncertainty} results in a statistical uncertainty in the measured phase of $\SI{92.7}{ppm}$.
Applying this to both the null and source mass measurements, the total statistical uncertainty is $\SI{131.1}{ppm}$.

\section{Conclusion}

The significant uncertainty in the CODATA 2022 value for G underscores the necessity that new meteorological devices capable of measuring gravity are designed and new measurements with systematic effects independent of previous experiments are conducted.
The mass and electric neutrality of neutrons make them a convenient probe for the relatively weak gravitational potentials created by laboratory-scale source masses.
The neutron PGMI, with its wide wavelength acceptance and large interferometer area, has overcome the previous limitations of single-wavelength NIs.
Here, we have shown that a neutron 3-PGMI has sufficient sensitivity to measure the phase shift induced by a $\SI{1}{\tonne}$ source mass and extract the gravitational constant with an uncertainty of $\SI{150}{ppm}$ [i.e., $G=\SI{6.674\pm0.001e-11}{\meter\cubed\per\kilo\gram\per\second\squared}$] with systematic effects distinct from state-of-the-art experiments.
A measurement of G to $\SI{100}{ppm}$ is possible with this design but would require more than $\SI{800}{\day}$ of experiment time in the current configuration.
Improvements offering greater neutron flux, smaller grating periods, or different source mass potentials may increase the gravitationally-induced phase shift and therefore improve the achievable precision in G with a neutron PGMI.

Without new methods of measuring gravitational forces, it is difficult to distinguish discrepancies in G caused by systematic effects from new physics~\cite{schlamminger2018gravity}, variations in Earth's gravitation~\cite{anderson2015measurements}, or relativistic effects~\cite{parra2018toward}.
Furthermore, this analysis of the systematic effects in a neutron 3-PGMI suggests that existing experiments measuring G may require a correction factor to account for time-dependent changes in tidal forces.
These corrections depend on both the experiment date and location, possibly accounting for some of the variation between experimentally measured values of G.
The model introduced here can also be used to design new operational modalities capable of measuring internal magnetic structures, like in skyrmion materials, or fundamental properties, like the neutron electric dipole moment.

\section{Acknowledgments}

This work was supported by the Canadian Excellence Research Chairs (CERC) program, the Natural Sciences and Engineering Research Council of Canada (NSERC) Discovery program, the Collaborative Research and Training Experience (CREATE) program, and the Canada  First  Research  Excellence  Fund  (CFREF).
We acknowledge the support of the National Institute of Standards and Technology, U.S. Department of Commerce, and US Department of Energy, Office of Nuclear Physics, under Interagency Agreement 89243019SSC000025.

\bibliographystyle{IEEEtran}
\bibliography{ref}

\begin{thebibliography}{10}
\bibitem{NIST_disclaimer}
Certain trade names and company products are mentioned in the text or
  identified in an illustration in order to adequately specify the experimental
  procedure and equipment used. In no case does such identification imply
  recommendation or endorsement by the National Institute of Standards and
  Technology, nor does it imply that the products are necessarily the best
  available for the purpose.
\end{thebibliography}

\clearpage

\appendix

\section{Model Validation}
\label{sec:model-validation}

\begin{figure}[h]
    \centering
    \includegraphics[width=\linewidth]{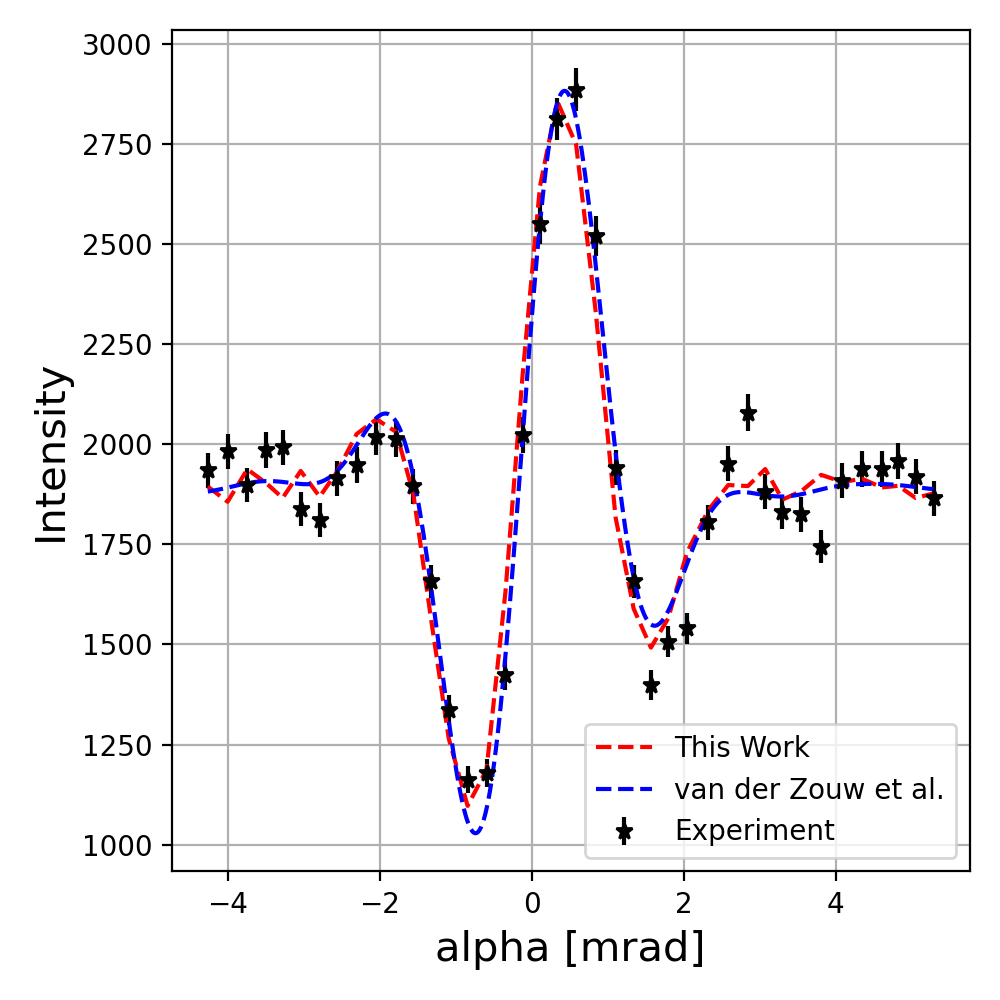}
    \caption{Neutron intensity as a function of grating rotation (alpha) from van der Zouw et al.~\cite{vanderZouw2000AharonovBohmInterferometer}.
    Experimental neutron counts are shown as black points, with the theoretical curve of Ref.~\cite{vanderZouw2000AharonovBohmInterferometer} shown as a blue dotted line.
    Monte Carlo simulation results from this work are shown as a red dotted line.
    }
    \label{fig:zeilinger}
\end{figure}

The Monte Carlo model described in this work, with example scripts used to compare simulation to neutron experiments, can be found at \url{https://git.uwaterloo.ca/c3kapahi/neutron-3pgmi-big-g-model-public}.
Several experiments have investigated the interference pattern generated by a series of phase gratings in a neutron beam, some including the effect of gravitational potentials.
In 2000, van der Zouw et al. measured the intensity of one path pair in a phase grating NI, namely the interference between the $\{0, 1, -1\}$ and $\{1, -1, 0\}$ orders, where each integer in the set denote the diffraction order from the $\{1^{\text{st}}, 2^{\text{nd}}, 3^{\text{rd}}\}$ grating~\cite{vanderZouw2000AharonovBohmInterferometer}.
The wavevector of a neutron within this interferometer will depend on its position, $\mathbf{r}$, as

\begin{equation}
    \mathbf{k}^2 = \mathbf{k}_0^2 + m_N g \mathbf{r}\cdot\mathbf{\hat{y}},
\label{eqn:cow-wavevector}
\end{equation}

\noindent
where $\mathbf{r} = 0$ is the coordinate of the entrance slit, $\mathbf{k}_0$ is the neutron's initial wavevector, $m_N$ is the neutron mass, and $g$ is the local acceleration due to gravity.
The interferometer in Ref.~\cite{vanderZouw2000AharonovBohmInterferometer} was simulated using the listed total interferometer length of $\SI{0.76}{\meter}$ with grating periods of $\SI{2}{\micro\meter}$ and the wavelength distribution shown in Figure 1 of that work.
The result of the path integral model of this experiment is shown as a red dotted line in Figure~\ref{fig:zeilinger}, with the experimental counts shown as black dots.
The path integral model (red) shows good agreement with the theoretical curve (blue), shown in Figure~\ref{fig:zeilinger} of Ref.~\cite{vanderZouw2000AharonovBohmInterferometer}.

Unlike the phase grating interferometer used by van der Zouw, the neutron PGMI generates an interference pattern from a range of incident neutron angles with no restriction on the allowed diffraction orders.
Experimental demonstrations of the neutron PGMI have used both two- and three-grating arrangements.
Experiments using two gratings have shown high contrast with a variety of neutron spectral distributions~\cite{Pushin2017Far-fieldImaging, sarenac2023cone}, while the three-grating arrangement has only been measured with low contrast~\cite{Sarenac2018ThreeApplications}, possibly due to misalignment of the phase gratings~\cite{Heacock2019AngularVisibility, sarenac2023cone}.

\begin{figure*}
    \centering
    \includegraphics[width=0.45\textwidth]{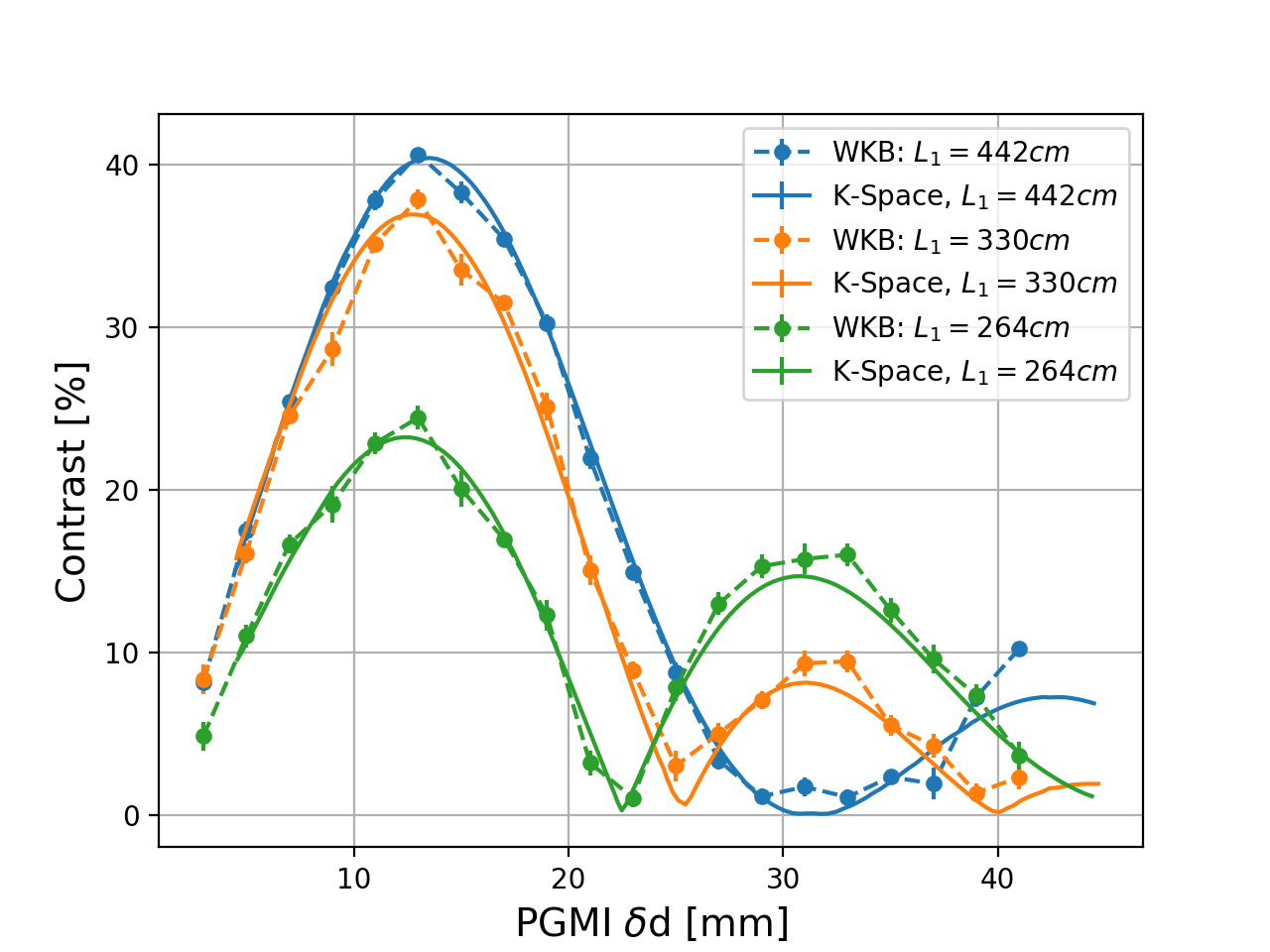}
    \includegraphics[width=0.45\textwidth]{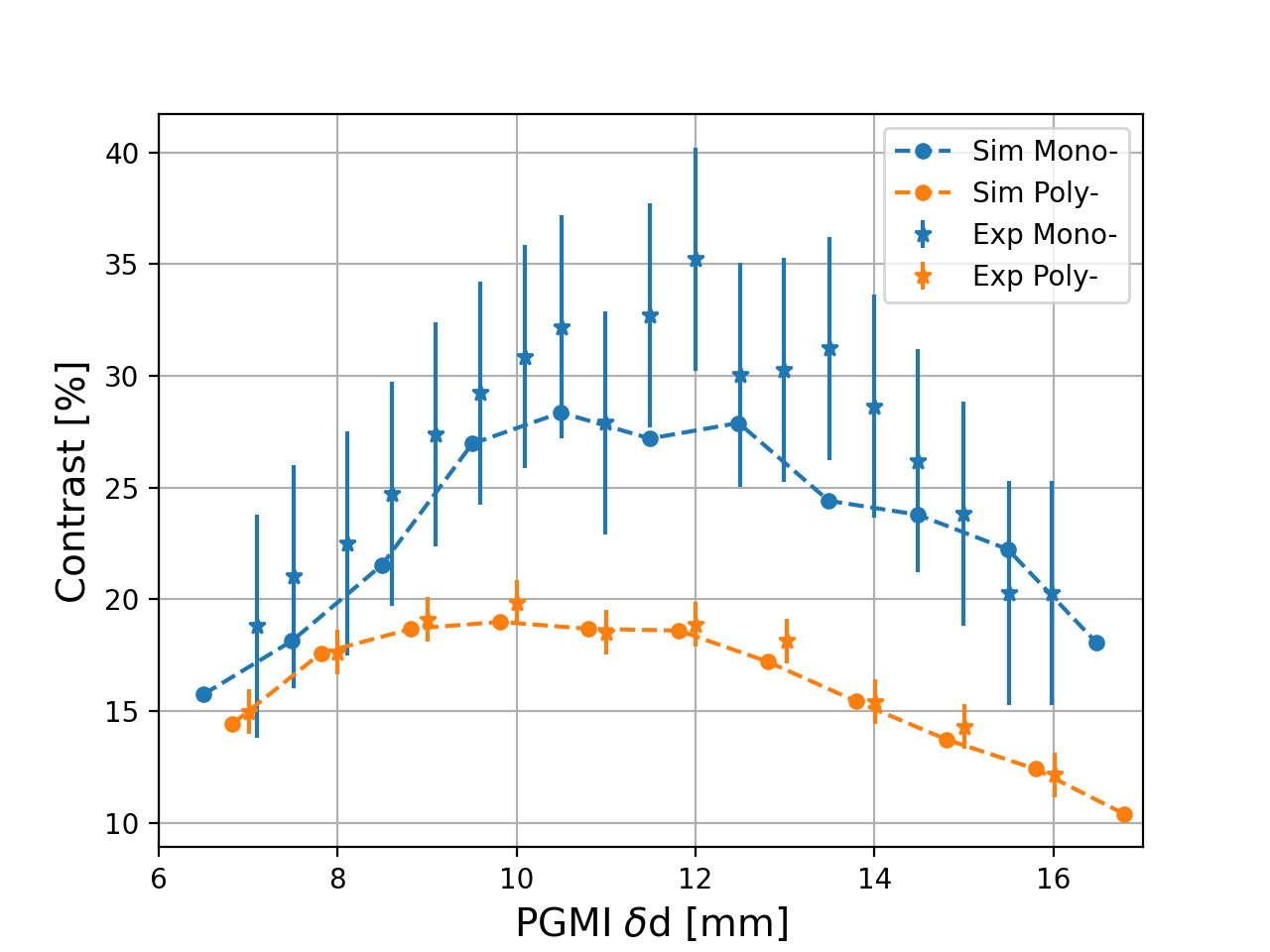}
    \caption{(left) Simulated contrast values for two-grating neutron PGMI experiment.
    Entrance slit to first grating distance, $L_1$, varies between datasets.
    Experimental parameters and original data can be found in Ref.~\cite{sarenac2023cone}.
    Simulated contrasts with the path integral model described in this work show good agreement with the plane-wave model in Ref.~\cite{sarenac2023cone}.
    (right) Simulated and experimental contrast values for two-grating neutron PGMI.
    Experimental parameters and original data can be found in Ref~\cite{Pushin2017Far-fieldImaging}.
    }
    \label{fig:jparc-nist}
\end{figure*}

Figure~\ref{fig:jparc-nist} shows the contrast of interference patterns generated by a Monte Carlo simulation of Eqn.~\ref{eqn:path-amplitude} with varying grating separation, $\delta d$.
Simulations with the WKB model described in this work are in good agreement with the momentum space model of ~\cite{sarenac2023cone} and with both the monochromatic and polychromatic PGMI experiments of ~\cite{Pushin2017Far-fieldImaging}.

\section{Gravitational Potential of a Right Rectangular Prism}
\label{sec:prism-potential}

The gravitational potential of any source mass can be computed numerically; however, analytic solutions exist for the potential from several simple shapes, including the right rectangular prism~\cite{nagy2000gravitational, karcol2019two}.
For a rectangular prism source mass of dimension $\{L_x, L_y, L_z\}$, the gravitational potential at a point, $(x, y, z)$, outside the mass volume is

\begin{widetext}
\begin{align}
    U_g(x, y, z) &= G\rho \int^{L_z/2}_{-L_z/2} \int^{L_y/2}_{-L_y/2} \int^{L_x/2}_{-L_x/2} \frac{dXdYdZ}{\sqrt{(x-X)^2 + (y-Y)^2 + (z-Z)^2}} \\
    &= \Bigg| \Bigg| \Bigg| XY\log(Z + |\mathbf{R}|) + YZ\log(X + |\mathbf{R}|) + ZX\log(Y + |\mathbf{R}|) \nonumber\\
    &\hspace{1cm} -\frac{X^2}{2}\tan^{-1}\bigg(\frac{YZ}{X|\mathbf{R}|}\bigg) -\frac{Y^2}{2}\tan^{-1}\bigg(\frac{ZX}{Y|\mathbf{R}|}\bigg) -\frac{Z^2}{2}\tan^{-1}\bigg(\frac{XY}{Z|\mathbf{R}|}\bigg) \Bigg|^{\sfrac{L_x}{2}-x}_{X = -\sfrac{L_x}{2}-x} \Bigg|^{\sfrac{L_y}{2}-y}_{Y = -\sfrac{L_y}{2}-y} \Bigg|^{\sfrac{L_z}{2}-z}_{Z = -\sfrac{L_z}{2}-z},
\end{align}
\end{widetext}

\noindent
where the origin of both coordinates $(x, y, z)$ and $(X, Y, Z)$ are the geometric center of the rectangular prism source mass, the density of the source mass, $\rho$, is uniform, and $\mathbf{R}$ is the distance from the source mass element at $(X, Y, Z)$ to the point of interest $(x, y, z)$.

\section{Phase Fit Statistics}
\label{sec:fit-stats}

This analysis expands on the work in Chapter 4.4 of Ref.~\cite{Werner2001NeutronMechanics}.
The underlying assumption in this analysis is that the number of counts, $N_j$, in a detector or single-pixel over some time, $T$, obeys a Poisson distribution, with a mean count across the experiment of $\Bar{N}$, and standard deviation $\sigma_{N_j} = \sqrt{\Bar{N}}$.
For our interferometer, the expected number of particles per pixel/detector position, $N_j$, is

\begin{equation}
    N_j = \Bar{N}\left[ 1 + \mathcal{C} \cos\left( \frac{2\pi}{P_f}x_j + \Delta\Phi_j \right) \right],
\label{eqn:neutron-count}
\end{equation}

\noindent
where $\mathcal{C}$ is the contrast of these interference fringes, $P_f$ if the period of the interference fringes, $x_j$ is the position of pixel $j$, and $\Delta\Phi_j$ is the phase difference between interfering paths as measured at this detector.

The variances in phase and detector count are related by

\begin{equation}
    (\sigma_{N_j})^2 = \left( \frac{\partial N_j}{\partial \Delta\Phi_j} \right)^2 (\sigma_{\Delta\Phi_j})^2,
\end{equation}

\noindent
where the partial derivative can be computed from Eqn.~\ref{eqn:neutron-count},

\begin{align}
    \left(\frac{\partial N_j}{\partial \Delta\Phi_j} \right)^2 &= \Bar{N}^2 \mathcal{C}^2 \sin^2\left( \frac{2\pi}{P_f}x_j + \Delta\Phi_j \right) \nonumber\\
    &= \Bar{N}^2 \mathcal{C}^2 - (N_j - \Bar{N})^2 \nonumber\\
    \implies (\sigma_{\Delta\Phi_j})^2 &= \frac{N_j}{\Bar{N}^2(\mathcal{C}^2 - 1) + 2\Bar{N}N_j - N_j^2},
\label{eqn:pixel-phase-uncert}
\end{align}

\noindent
where the relation, $\sin^2x + \cos^2x = 1$, has been used along with Eqn.~\ref{eqn:neutron-count}.
The total uncertainty in $\Delta\Phi$ from fitting all pixels across the detector follows from a $\chi^2$ optimization over $M$ different ``detectors'' or over $M$ pixels,

\begin{equation}
    (\sigma_{\Delta\Phi})^2 = \left[ \sum_{j=1}^M (\sigma_{\Delta\Phi_j})^{-2} \right]^{-1}.
\label{eqn:phase-sum}
\end{equation}

\noindent
The sum over $M$ pixels can be approximated using an integral over a finite number of periods.
Choosing to approximate this sum as an integral over one period, $P$, the sum in Eqn.~\ref{eqn:phase-sum} is

\begin{align}
    \sum_{j=1}^M (\sigma_{\Delta\Phi_j})^{-2} \approx \frac{1}{2\pi}&\int_0^{2\pi} \frac{\Bar{N}^2(\mathcal{C}^2 - 1) + 2\Bar{N}N_j - N_j^2}{N_j} dx \nonumber\\
    &\approx \frac{\Bar{N} (\mathcal{C}^2 - 1)}{\sqrt{1 - \mathcal{C}^2}} + 2\Bar{N} - \Bar{N},
\label{eqn:phase-sum-eval}
\end{align}

\noindent
where the average counts, $\Bar{N}$, applies across the range of pixels used in the experiment.
Using the relationship, $\Bar{N} = \dot{N}T$, where $\dot{N}$ is the count rate at the detector and $T$ is the experiment duration, we can combine Eqn.~\ref{eqn:phase-sum} and ~\ref{eqn:phase-sum-eval} to find

\begin{equation}
    (\sigma_{\Delta\Phi})^2 \approx \frac{1}{\dot{N}T\left(1 - \sqrt{1-\mathcal{C}^2} \right)},
\label{eqn:apdx-phase-uncertainty}
\end{equation}

\noindent
as shown in Section~\ref{sec:stats}.
We can approximate the square-root term in Eqn.~\ref{eqn:apdx-phase-uncertainty} with $\mathcal{C}^2 << 1$ and rearrange these terms to estimate the required beam time for a measurement as

\begin{equation}
    T \approx \frac{2}{\dot{N}(\sigma_{\Delta\Phi}\cdot\mathcal{C})^2},
\end{equation}

\noindent
which suggests that minimizing the required beam time for a measurement is achieved by maximizing the product $(\sigma_{\Delta\Phi}\cdot\mathcal{C})$, as stated in Section~\ref{sec:proposal}.

\end{document}